\documentclass[%
reprint,  
amsmath,amssymb,
aps,
prd,
]{revtex4-2}

\usepackage{graphicx}
\usepackage{ctable}        
\usepackage{array}         
\usepackage{xcolor}        
  \definecolor{navy}{HTML}{2c1fa5}
  \definecolor{maroon}{HTML}{800000}
  \definecolor{gray}{HTML}{808080}
\usepackage[final]{hyperref} 
\hypersetup{
	colorlinks=true,         
  linkcolor=navy,          
	citecolor=navy,          
	filecolor=magenta,       
	urlcolor=navy         
}

\usepackage{texmacros}     

\begin{document}

\title{Dynamics of $U(1)$ gauged \qballs{} in three spatial dimensions}

\author{Michael P. Kinach}
 \email{mikin@physics.ubc.ca}
\author{Matthew W. Choptuik}%
 \email{choptuik@physics.ubc.ca}
\affiliation{%
  Department of Physics and Astronomy, University of British Columbia,\\
  6224 Agricultural Road, Vancouver, British Columbia, V6T 1Z1, Canada
}%

\date{\today}

\begin{abstract}
  We investigate the dynamics of $U(1)$ gauged \qballs{} using fully
  three-dimensional numerical simulations. We consider two different scenarios:
  first, the classical stability of gauged \qballs{} with respect to generic
  three-dimensional perturbations, and second, the behaviour of gauged
  \qballs{} during head-on and off-axis collisions at relativistic velocities.
  With regard to stability, we find that there exist gauged \qball{}
  configurations which are classically stable in both logarithmic and
  polynomial scalar field models.  With regard to relativistic collisions, we
  find that the dynamics can depend on many different parameters such as the
  collision velocity, relative phase, relative charge, and impact parameter of
  the colliding \qballs{}. 
\end{abstract}

\maketitle


\section{\label{sec:intro} Introduction }

\qballs{} are non-topological solitons that arise in scalar field theories
admitting a $U(1)$ symmetry and a non-linear attractive potential. First
described by Coleman \cite{Coleman1985,*Coleman1986}, they have garnered
significant attention in recent years due to their potential relevance to
early-Universe cosmology (where they may act as dark matter candidates
\cite{Kusenko1998,Kusenko2001}) and in condensed matter experiments (where they
serve as relativistic analogues to various condensed matter solitons
\cite{Enqvist2003,Bunkov2007,Autti2018}). \qballs{} also hold considerable theoretical
interest as smooth, classical field configurations which constitute a
rudimentary model of a particle.

An extension to the basic \qball{} theory can be made through the introduction
of a $U(1)$ gauge field. This gives rise to so-called \textit{gauged \qballs{}}
which couple to the electromagnetic field and carry an electric charge
\cite{Lee1989}. While gauged \qballs{} share some similarities with ordinary
(non-gauged) \qballs{}, the additional electromagnetic coupling can also lead
to several distinct features.  For example, it may place restrictions on their
allowable size and charge \cite{Tamaki2014,Gulamov2015}, change their dynamical
behaviour \cite{Kinach2023,Kinach2024}, and even give rise to new types of
solutions in the model (such as shell-shaped structures
\cite{Arodz2009,Tamaki2014,Heeck2021shell}). It has also been speculated that the
repulsive Coulomb force arising from a gauged \qball{} might serve as a
destabilizing mechanism which eventually destroys it \cite{Panin2017}. This is
an important issue because one should expect gauged \qballs{} to be robust
against generic perturbations in order to be considered viable physical
objects. However, the stability analysis of these objects is challenging 
because the application of
standard methods for establishing classical stability (such as linear
perturbation analyses or known stability theorems) are hindered by the presence
of the $U(1)$ gauge field.
In particular, it is known that gauged \qballs{}
can be classically stable against spherically-symmetric and axially-symmetric
perturbations \cite{Panin2017,Kinach2023}, but the case of general
three-dimensional perturbations has yet to be explored.

In the present work, we address this problem of gauged \qball{} stability by
performing fully non-linear numerical evolutions of the equations of motion in
three spatial dimensions. For gauged \qballs{} in both logarithmic and
polynomial scalar field models, we find numerical evidence for solutions which
are classically stable against generic three-dimensional perturbations over
long dynamical timescales. In these cases, we find that the stable gauged
\qballs{} respond to the perturbations by oscillating continuously or weakly
radiating before evolving toward a state that is close to the initial configuration. In other cases,
we also observe examples of unstable configurations which are eventually 
destroyed by the perturbations (for instance, by fragmentation into smaller
gauged \qballs{}).
Our results are found to be generally consistent with previous
numerical work on gauged \qball{} stability under spherical and axial symmetry
assumptions \cite{Panin2017,Kinach2023}. Motivated by the
very recent analysis of \cite{Rajaraman2024}, we also investigate the
case of the polynomial scalar field potential at small gauge coupling
and find a new result for the instability transition point in comparison
to what was reported in \cite{Kinach2023}.

Another question we explore relates to the behaviour of gauged \qballs{}
during relativistic collisions.  In \cite{Kinach2024}, it was shown that gauged
\qballs{} can exhibit a range of remarkable interaction phenomena such as
mergers, fragmentation, charge transfer, charge annihilation, \qring{}
formation, and radiation production. However, these results have also been
limited by the assumption of axial symmetry. It is worthwhile to ask whether
any of these phenomena are peculiar to axial symmetry or whether they also
extend to a more realistic three-dimensional setting. Moreover, it is
interesting to ask how the dynamics may change during gauged \qball{}
collisions with non-zero impact parameter (a scenario which was not accessible
under previous symmetry assumptions). In the present work, we address these
questions by considering both head-on and off-axis collisions of gauged
\qballs{} in three spatial dimensions.

This paper is organized as follows: in Sec.~\ref{sec:eom}, we present the basic
equations of the theory. In Sec.~\ref{sec:implement}, we describe our numerical
implementation of the evolution equations along with our initial data
procedure.  In Sec.~\ref{sec:results}, we present our main numerical results.
In Sec.~\ref{sec:conclusion}, we provide some concluding remarks.

Throughout this work, we employ units where $c=\hbar=1$.
For brevity, we interchangeably use the terms ``\qball{}" and ``gauged
\qball{}" when the distinction between the gauged and non-gauged solutions is
made obvious by context. 

\section{\label{sec:eom} Equations of Motion }

The theory of $U(1)$ gauged \qballs{} can be described by the Lagrangian density
\begin{equation}
  \mathcal{L} = -\left(D_\mu\phi\right)^* D^\mu\phi-V\left(|\phi|\right)-\frac{1}{4}F_{\mu\nu}F^{\mu\nu}.
  \label{eqn:gauged-lagr}
\end{equation}
Here, $\phi$ is the complex scalar field, $A_\mu$ is the $U(1)$ gauge field,
$D_\mu = \nabla_\mu-ieA_\mu$ is the gauge covariant derivative with coupling
constant $e$, $F_{\mu\nu}=\partial_{\mu} A_\nu - \partial_{\nu} A_\mu$ is the
electromagnetic field tensor, and $V(|\phi|)$ is the scalar potential.  The
equations of motion for the theory take the form
\begin{align}
  D_\mu D^\mu \phi - \frac{\partial}{\partial \phi^*}V(|\phi|)&=0, \label{eqn:eom-a}\\
  \nabla_\mu F^{\mu\nu}+ej^\nu&=0, \label{eqn:eom-b}
\end{align}
where 
\begin{equation} \label{eqn:j-current}
  j^\nu=-i(\phi^* D^\nu\phi-\phi(D^\nu\phi)^*)
\end{equation}
is the Noether current density.  Consistent with previous work
\cite{Kinach2023,Kinach2024}, we consider two forms for the scalar field
potential:
\begin{align}
  V_\text{log}(|\phi|)&=-\mu^2|\phi|^2\ln(\beta^2|\phi|^2),\label{eqn:log}\\
  V_\text{6}(|\phi|)&=m^2|\phi|^2-\frac{k}{2}|\phi|^4+\frac{h}{3}|\phi|^6,\label{eqn:poly}
\end{align}
where $\mu$, $\beta$, $m$, $k$, and $h$ are real, positive parameters.
Additionally, we employ the Minkowski line element,
\begin{equation}
  ds^2=-dt^2+dx^2+dy^2+dz^2,
  \label{eqn:minkowski}
\end{equation}
and fix the gauge with the Lorenz condition,
\begin{equation}
  \label{eqn:lorenz}
  \nabla_\mu A^\mu=0,
\end{equation}
in order to write the equations of motion
\eqref{eqn:eom-a}--\eqref{eqn:eom-b} in a form which is suitable for
numerical evolution (see App.~\ref{sec:evoeqns}). In addition to these
equations, solutions in the theory \eqref{eqn:gauged-lagr} must also
satisfy the constraints
\begin{align}
  \nabla_i E^i&=ej^0 \label{eqn:divE},\\
  \nabla_i B^i&=0 \label{eqn:divB}.
\end{align}
Here, $E^i$ and $B^i$ represent the components of the electric and
magnetic field vectors, respectively, which are determined from the
electromagnetic field tensor, $F_{\mu\nu}$.  Solutions in the theory
\eqref{eqn:gauged-lagr} are expected to satisfy
\eqref{eqn:divE}--\eqref{eqn:divB} everywhere in the solution domain.
The amount by which these constraints are
violated therefore provides a relative measure of the error in the
numerical evolution; this issue will be discussed in further detail
below.

\section{\label{sec:implement} Numerical Implementation }

As stated previously, we use a numerical framework to study the dynamics
of the model in three spatial dimensions. Here we provide the details of
this approach.

\subsection{Initial Data}
\label{sec:initdata}

In order to generate initial data which describes gauged \qballs{},
we begin by making a spherically-symmetric ansatz for the fields,
\begin{align}
  \phi(t,\vec{x})&=f(r)e^{i\omega t},  \label{eqn:sph-ansatz-f} \\
  A_0(t,\vec{x})&=A_0(r), \label{eqn:sph-ansatz-A} \\
  A_i(t,\vec{x})&=0.  \label{eqn:sph-ansatz-B}
\end{align}
With this ansatz, the equations of motion reduce to a system of two
coupled differential equations,
\begin{align}
  f''(r)+\frac{2}{r}f'(r)+f(r)g(r)^2-\frac{1}{2}\frac{d}{df}V(f)&=0,
  \label{eqn:shooting-a}\\
  A_0''(r)+\frac{2}{r}A_0'(r)+2ef(r)^2g(r)&=0,
  \label{eqn:shooting-b}
\end{align}
where we have defined $g(r)=\omega-eA_0(r)$. To find gauged \qball{} solutions
which are smooth with finite energy, we impose the boundary conditions:
\begin{alignat}{2}
  \frac{df}{dr}(0)=0,\qquad\quad &&\lim_{r\rightarrow \infty}f(r)=0,
  \label{eqn:shooting-bc-a}\\
  \frac{dA_0}{dr}(0)=0,\qquad\quad && \lim_{r\rightarrow \infty}A_0(r)=0.
  \label{eqn:shooting-bc-b}
\end{alignat}
Together, the differential system \eqref{eqn:shooting-a}--\eqref{eqn:shooting-bc-b}
is akin to
an eigenvalue problem for the parameter $\omega$. As described in
\cite{Kinach2023}, we use a numerical shooting technique to solve this system for 
$f(r)$ and $A_0(r)$ to a good approximation. The resultant solutions
provide the spherically-symmetric profile functions for gauged \qballs{} at
a given value of $\omega$.

To initialize the fields in three dimensions, it is necessary to compute
the values of the spherical functions $f(r)$ and $A_0(r)$ at arbitrary
points in space using the coordinate system defined by
\eqref{eqn:minkowski}. For this purpose, we apply fourth-order Neville
interpolation \cite{Press2007} to the numerical profiles of $f(r)$ and
$A_0(r)$ and set the values of $\phi$ and $A_\mu$ using the ansatz
\eqref{eqn:sph-ansatz-f}--\eqref{eqn:sph-ansatz-B}. With this procedure,
it is straightforward to construct the initial data for a single
stationary gauged \qball{} which is centered at the origin. This is the
form of initial data we use to study gauged \qball{} stability.

When studying relativistic collisions of gauged \qballs{}, the
previously-described procedure must be adjusted. The main difference
comes from the need to initialize a binary configuration of \qballs{}
which are Lorentz-boosted at a relativistic velocity $v$ (where $v=1$ is
the speed of light in our units). In this case, an initial displacement
from the origin is chosen for each \qball{} and the Neville
interpolation procedure is performed separately about the
center point for each soliton. Each
gauged \qball{} is then given a Lorentz boost
in a direction parallel to the $z$-axis and toward the origin.  Finally,
the fields of each gauged \qball{} are superposed to complete the
initial data specification.

As discussed in \cite{Kinach2024}, some care must be taken when
implementing the above procedure for binary gauged \qballs{}. In
particular, if the \qball{}s in the binary are not sufficiently
separated at the initial time, the long-range behaviour of the gauge
field can lead to unphysical violations of the constraint equation
\eqref{eqn:divE}. These arise due to the influence of the gauge field
from one \qball{} on the scalar field of the other. 
In an ideal case, one could avoid this problem by picking a sufficiently large
separation distance so that these influences are negligible. However,
this proves to be impractical for our numerical simulations because
large initial separation distances incur a greater computational cost.
Instead, we address this problem by implementing an FAS multigrid
algorithm with fourth-order defect correction \cite{Trottenberg2000} to
re-solve the constraint equation \eqref{eqn:divE} at the initial time
for general superpositions of gauged \qball{}s (see also \cite{York1982}). This provides an
order-of-magnitude reduction in the constraint violation associated with
our binary initial data.

\subsection{Diagnostic Quantities}

Here we describe a number of diagnostic quantities which can be used to
assess the numerical results. Foremost among these are the total energy $E$
and total Noether charge $Q$ which are conserved in the continuum limit.
For the theory described by
\eqref{eqn:gauged-lagr}, the energy-momentum tensor takes the form
\begin{equation} \label{eqn:emt}
  \begin{split}
    T_{\mu\nu} =\, &F_{\mu\alpha} F_{\nu\beta} g^{\beta\alpha} -\frac{1}{4}g_{\mu\nu}F_{\alpha\beta}F^{\alpha\beta}\\
  &+D_\mu\phi(D_\nu\phi)^*+D_\nu\phi(D_\mu\phi)^*\\
  &-g_{\mu\nu}(D_\alpha\phi(D^\alpha\phi)^*+V(|\phi|)).
  \end{split}
\end{equation}
Using \eqref{eqn:emt}, we define the total energy contained in the system as
$E=\int T_{00}\,d^3x$. Likewise, the total Noether charge can be computed from
the current density \eqref{eqn:j-current} as $Q=\int j^{0}\,d^3x$.  In all
simulations discussed below, these quantities are monitored to ensure that they
do not deviate from their initial values by more than $O(1\%)$.

In order to investigate the dynamical stability of gauged \qballs{}, it is necessary
to introduce small perturbations into the system. For this purpose, we incorporate an
auxiliary scalar field into the theory \eqref{eqn:gauged-lagr} which serves
as a diagnostic tool. The modified Lagrangian density of the theory
takes the following form:
\begin{equation} \label{eqn:gauged-lagr-pert}
\begin{split}
  \mathcal{L} = -\left(D_\mu\phi\right)^* D^\mu\phi &-
  V\left(|\phi|\right)-\frac{1}{4}F_{\mu\nu}F^{\mu\nu}\\&-\partial_\mu\chi\partial^\mu\chi
  - U(|\phi|,\chi).
\end{split}
\end{equation}
Here, $\chi$ is a massless real scalar field which couples to the complex
\qball{} field $\phi$ via the interaction potential $U(|\phi|,\chi)$.  As
discussed in \cite{Kinach2023}, the auxiliary field $\chi$ can act as an external perturbing
agent if the initial data and interaction potential $U(|\phi|,\chi)$ are chosen
so that $\chi$ exerts a small, temporary influence on
$\phi$. In particular, if $\chi$ is chosen to take the form of an aspherical
pulse which implodes onto a stationary gauged \qball{} at the origin, the
interaction governed by $U(|\phi|,\chi)$ is expected to excite all underlying
modes of the configuration. If the configuration is stable, we expect the
oscillations of these modes to remain bounded and the \qball{} to
stay intact. However, if the configuration is unstable,
we expect that one or more modes will grow exponentially,
eventually bringing about the destruction of the gauged \qball{} in
some manner (for example, via fragmentation or dispersal of the fields).
In this way, we can probe the stability properties of gauged \qballs{}
by observing their interaction with the auxiliary field $\chi$. 

Here we choose the scalar interaction potential in 
\eqref{eqn:gauged-lagr-pert} to take the form
\begin{equation}
  U(|\phi|,\chi)=\gamma|\phi|^2\chi^2
  \label{eqn:U-pert}
\end{equation}
and initialize the perturbing field according to
\begin{equation}
   \chi(0,x,y,z)=A\exp\left[-\left(\frac{\Delta-r_0}{\delta}\right)^2\right]
   \label{eqn:chi}
\end{equation}
where
\begin{equation}
  \Delta=\sqrt{\frac{(x-x_0)^2}{a_x^2}+\frac{(y-y_0)^2}{a_y^2}+\frac{(z-z_0)^2}{a_z^2}}.
  \label{eqn:Delta}
\end{equation}
In the above, $A$, $\delta$, $r_0$, $a_x$, $a_y$, $a_z$, $x_0$, $y_0$ and $z_0$ are
real, positive parameters which determine the initial profile of $\chi$.
In particular, if $r_0$ is large, then \eqref{eqn:chi} resembles a
shell-like concentration of the field which approximately vanishes in
the vicinity of the \qball{} at $t=0$. This shell can be made to implode
upon the origin at some time $t>0$ by setting
\begin{equation}
   \partial_t\chi(0,x,y,z)=\frac{\chi+x\partial_x\chi+y\partial_y\chi+z\partial_z\chi}{\sqrt{x^2+y^2+z^2}}.
\end{equation}
The form of the interaction potential
\eqref{eqn:U-pert} means that, after implosion, 
$\chi$ will propagate out toward
infinity at late times, leaving no significant remnant near the origin.
Thus, $\chi$ represents a time-dependent perturbation whose
influence on the \qball{} field $\phi$ can be directly controlled via
the parameter $A$ in \eqref{eqn:chi} (or similarly, via $\gamma$ in
\eqref{eqn:U-pert}).

While the auxiliary field $\chi$ serves as a convenient diagnostic tool
for our purposes, we emphasize that it is by no means the only form of
perturbation which exists in the system. 
In particular, our finite-difference approach
for solving the equations of motion (to be described below) inherently
introduces small-scale errors into our simulations which also act as
perturbations.  However, 
given the nature of the finite-difference scheme we use, as well as the 
typical numerical resolution we adopt, this type of perturbation is typically
very small;
this can make it
difficult to definitively assess the stability of the \qball{} unless
the simulation timescale is very long. By introducing the
field $\chi$ in \eqref{eqn:gauged-lagr-pert}, we gain an additional
level of control over the perturbative dynamics of the system beyond
what is possible in the original (unmodified) theory
\eqref{eqn:gauged-lagr}. 

\subsection{Evolution Scheme}
\label{sec:evoscheme}

To solve the equations of motion of the system in three spatial
dimensions, we use a fourth-order 
finite-difference scheme implemented using the Finite Difference Toolkit
(FD) \cite{Akbarian2015}. A fourth-order classic Runge-Kutta method
\cite{Press2007} is used for the time integration. 
Additionally, a sixth-order Kreiss-Oliger dissipation operator is added to the
equations of motion in order to reduce deleterious effects of grid-scale
solution components arising from the finite-difference computations. We also
utilize a modified Berger-Oliger adaptive mesh refinement (AMR) algorithm
\cite{Pretorius2006} in order to tailor the numerical resolution of our
simulations according to local truncation error estimates.
We discuss the validation of our numerical code in App.~\ref{sec:codevalid}.

As in \cite{Kinach2024}, we find it advantageous when solving the
equations of motion to invoke a change of coordinates
$x^\mu=(t,x,y,z)\rightarrow x^{\mu'}=(t,X,Y,Z)$ according to
\begin{align}
  x &= d\exp(cX) - d\exp(-cX), \label{eqn:compactX} \\
  y &= d\exp(cY) - d\exp(-cY), \label{eqn:compactY} \\
  z &= d\exp(cZ) - d\exp(-cZ), \label{eqn:compactZ}
\end{align}
where $c$ and $d$ are positive, real parameters.  With the
transformations defined by \eqref{eqn:compactX}--\eqref{eqn:compactZ},
the simulation domain can be approximately
compactified at large coordinate values
while retaining coordinates near the origin that are close to their 
untransformed values. This transformation is
advantageous for two reasons.  First, it allows us to observe the
dynamics in scenarios where appreciable field content may propagate
swiftly away from the origin and reach large coordinate distances. 
Second, it greatly simplifies the process of setting appropriate
boundary conditions for the problem. In particular, our fourth-order 
finite-difference scheme requires a spatial stencil which spans at least
five grid points in each spatial dimension (or seven grid points when applying sixth-order Kreiss-Oliger dissipation).
While this is
straightforward to implement in the interior of the domain, the boundary
regions (and surrounding area) require a meticulous treatment in terms
of fourth-order backward and forward difference operators.
However, with the coordinate transformations defined by
\eqref{eqn:compactX}--\eqref{eqn:compactZ}, the simulation domain can be
made large enough so that Dirichlet conditions can be imposed
as a reasonable approximation at the physical boundaries and at
boundary-adjacent points. This greatly reduces the complexity of the
implementation.

For all results presented in this work, we set a base-level grid
resolution of $129^3$ points and utilize up to 8 levels of additional
mesh refinement with a refinement ratio of 2:1. We select a Courant factor
of $\lambda = dt/ \{dX, dY, dZ\} = 0.25$ and choose $c=0.05$, $d=10$ in
the transformations \eqref{eqn:compactX}--\eqref{eqn:compactZ}. When
investigating the stability of gauged \qballs{}, we use a domain
$-150\le X,Y,Z \le 150$, corresponding to 
a physical domain given by approximately $-18000\le x,y,z \le 18000$.
When investigating
relativistic collisions of gauged \qballs{}, we use a domain with
$-75\le X, Y, Z \le 75$, corresponding to approximately
$-425\le x,y,z \le 425$. In both cases, the
Dirichlet boundary conditions imposed during the evolution are sampled
from the grid function values at the initial time. We have also verified
that these boundary conditions do not introduce any significant errors
which propagate inward and pollute the interior solution. 

\section{\label{sec:results} Numerical Results }

Here we present results from our numerical evolutions of 
the gauged \qball{} system. As stated above, we consider two
forms for the scalar potential (logarithmic \eqref{eqn:log} and polynomial \eqref{eqn:poly})
and set $\mu=\beta=m=k=1$ and $h=0.2$ following previous work \cite{Kinach2023,Kinach2024}. Due to the large computational
cost associated with fully three-dimensional evolutions, we restrict our
analysis to a few values of gauge couplings $e$. 
In particular, for the
logarithmic potential $V_\text{log}(|\phi|)$ in \eqref{eqn:log}, we examine
$e=1.1$, while for the polynomial potential $V_\text{6}(|\phi|)$ in
\eqref{eqn:poly}, we examine $e=0.17$ (which is near the maximum allowable
value for our choice of the potential parameters \cite{Loginov2020}) and $e=0.02$.
To illustrate some of the salient dynamics in these models, we will use three
specific gauged \qball{} configurations which are listed in Table
\ref{table:solns}.

{ \setlength\extrarowheight{2pt} \begin{table*}
  \begin{tabular}{|c|c|c|c|c|c|c|c|c|c|c|} \hline
    Configuration & Potential   & $e$     & $|\phi(0,0,0)|$       & $A_0(0,0,0)$  & $\omega$  & $E$      & $|Q|$    & Stable?  \\ \hline \hline
    A             & Logarithmic & $1.1$   & $0.6461$              & $1.383$       & $2.522$   & $52.08$  & $22.37$  & Yes      \\ 
    B             & Logarithmic & $1.1$   & $2.448\times10^{-13}$ & $0.9803$      & 3.078     & 260.3    & $92.76$  & No       \\ 
    C             & Polynomial  & $0.17$  & $1.973$               & $2.515$       & $0.9976$  & $405.1$  & $387.5$  & Yes      \\ 
    \hline
  \end{tabular}
  \caption{Table of representative gauged \qball{} configurations which
  are used to illustrate the dynamics in the theory
  \eqref{eqn:gauged-lagr}.  The configurations A and B correspond to the
  logarithmic potential \eqref{eqn:log}.  The configuration C
  corresponds to the polynomial potential \eqref{eqn:poly}.  From left
  to right, the remaining columns give the value of the
  electromagnetic coupling constant $e$, the initial central value of
  the scalar field $|\phi(0,0,0)|$, the initial central value of the
  gauge field $A_0(0,0,0)$, the \qball{} oscillation frequency $\omega$,
  the total energy $E$ of the solution (when stationary), and the total
  Noether charge $|Q|$ of the solution. The final column indicates the
  stability of the configuration as determined through our numerical
  simulations.}
\label{table:solns} \end{table*} }

While all calculations in this section are performed using the
compactified coordinates defined by
\eqref{eqn:compactX}--\eqref{eqn:compactZ}, we will hereafter present
all results using the original coordinates defined by the line element
\eqref{eqn:minkowski}. This is done mainly for ease of interpretation.

\subsection{Stability}
\label{sec:stability}

For the purposes of this work, we define the stability of a gauged
\qball{} configuration in terms of its response to small dynamical
perturbations.  Specifically, we consider a configuration to be stable
if physical quantities influenced by the perturbation---such as the field
maxima---remain bounded in time (aside from small numerical drifts which
may arise due to the long timescales used in our simulations). Unstable configurations, on the other
hand, are those for which some component of the fields may grow
continuously in response to the perturbation until the initial \qball{}
is destroyed.

As mentioned previously, we use an auxiliary real massless scalar field
$\chi$ as an external perturbing agent.  The field $\chi$ takes the form
of an imploding pulse which is slightly aspherical and off-center from
the origin at the initial time. This choice ensures that the gauged
\qball{} (which is initially centered at the origin) will experience a
generic three-dimensional perturbation which is likely to excite all
underlying modes of the solution. After the field $\chi$ explodes
through the origin, the subsequent behaviour of the \qball{} can be
observed. To make an assessment of stability, we compute the maximal
value of $|\phi|$ over the entire numerical domain.  If this maximal
value (which is presumed to be attained near the \qball{} center) oscillates
continuously near the initial value in response to the perturbation,
we conclude that the configuration is stable. We also visualize the
fields in 3D to observe whether there is any change in shape or
behaviour. If the field maximum or shape of the \qball{} significantly
and permanently deviates from the initial configuration (such as by
breaking apart into smaller structures), we conclude that the
configuration is unstable.

\begin{figure}
    \includegraphics[width=\columnwidth]{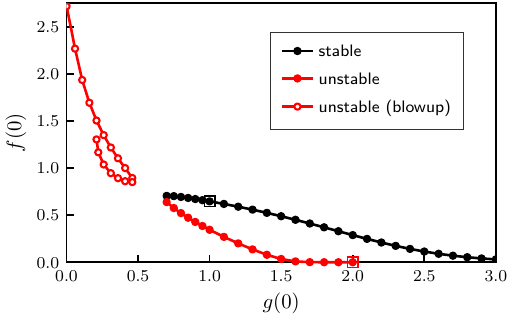}
    \caption{Shooting results and regions of stability and instability for
    gauged \qballs{} in the logarithmic model \eqref{eqn:log} with $e=1.1$.
    Plotted is the \qball{}'s central scalar field value $f(0)$ versus the
    numerical shooting parameter $g(0)=\omega-eA_0(0)$. The black solid circles
    represent configurations which are found to be stable with respect to
    generic three-dimensional perturbations. The red solid and open circles
    represent configurations which are found to be unstable with respect to
    these perturbations. The open squares represent configurations
    A and B from Table \ref{table:solns}.}
    \label{fig:solnspace-log}
\end{figure}

To begin the analysis, we use the shooting procedure described in
Sec.~\ref{sec:initdata} to obtain gauged \qball{} solutions for the
potentials \eqref{eqn:log} and \eqref{eqn:poly}. The space of solutions
for the logarithmic potential \eqref{eqn:log} with $e=1.1$ is depicted
in Fig.~\ref{fig:solnspace-log}.  In the figure, each dot represents one
distinct gauged \qball{} configuration which is found via the shooting
procedure. For each of these configurations, we evolve the system twice
to assess its stability. First, the evolution is performed with the
auxiliary field $\chi$ acting as an perturbing agent; for this we set
$\gamma=0.1$ in \eqref{eqn:U-pert} and $A=0.1$ in \eqref{eqn:chi} so
that the field has a material impact on the evolution of the \qball{}
field $\phi$.  Second, we perform the same evolution with $\gamma=0$ so
that $\chi$ and $\phi$ do not interact. In this case, the gauged
\qball{} is subject only to the small perturbations arising from the
truncation error of the scheme 
or other numerical sources (such as those associated with the AMR algorithm
\cite{Radia2022}). For both of these evolutions, we evolve the system
until at least $t=1200$ which typically corresponds to $O(100)$ internal
oscillations of the \qball{}. 
The outcome of the evolution is then classified depending on whether an
instability is observed. In Fig.~\ref{fig:solnspace-log}, the stable
configurations are marked by black solid circles while the unstable
configurations are marked by red solid and open circles.

By looking at Fig.~\ref{fig:solnspace-log}, one can observe several
interesting features. The first is the existence of both stable and
unstable branches in the space of gauged \qball{} solutions. By direct
comparison with previous work, one can see that the regions of stability
and instability correspond exactly with what has been found for
axisymmetric perturbations (cf.~Fig.~3 of \cite{Kinach2023}).
This suggests that three-dimensional perturbations do not excite any
additional unstable modes for gauged \qballs{} with $e=1.1$ in the
logarithmic model. The appearance of a stable branch also addresses the
general question of gauged \qball{} stability which was originally posed
in \cite{Panin2017} (namely, whether the Coulomb force will eventually
destroy any gauged \qball{} when symmetry assumptions are relaxed). This
reaffirms the possibility of gauged \qballs{} as viable physical objects
in realistic three-dimensional settings. 

\begin{figure}
    \includegraphics[width=\columnwidth]{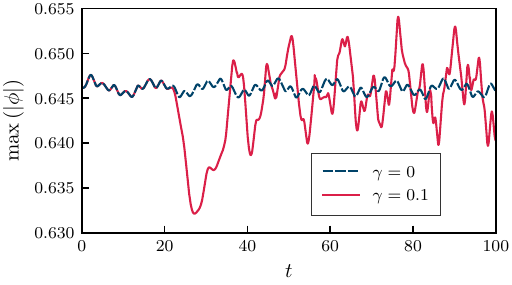}
    \caption{Oscillations in the maximum of the scalar field modulus
    $|\phi|$ for the stable gauged \qball{} corresponding to
    configuration A in Table \ref{table:solns}.  The results from two
    evolutions are shown. For the case where $\gamma=0$ in
    \eqref{eqn:U-pert}, the field $\chi$ has no influence on
    the evolution of the \qball{} and the fields are perturbed only by
    the inherent error of the numerical simulation.  For the case where
    $\gamma=0.1$, the field $\chi$ interacts with the \qball{} starting
    at $t\approx 20$ and induces relatively large oscillations in the \qball{}
    modulus.  We note that the amplitude of the induced oscillations for
    the case of $\gamma=0.1$ is highly dependent on the precise shape of
    the pulse as defined through \eqref{eqn:chi}--\eqref{eqn:Delta}.} 
    \label{fig:log-osc}
\end{figure}

Let us discuss in further detail the behaviour of these stable
configurations.  As previously stated, we perturb each configuration in
two ways: first, by the implosion of the field $\chi$, and second, by
truncation errors.  In both cases, we find that the
\qballs{} respond to the perturbations by oscillating continuously
around the equilibrium configurations. An illustration of this behaviour
is given in Fig.~\ref{fig:log-osc}. Initially, the gauged \qball{}
remains at the origin and is perturbed only by truncation error. At
$t\approx 20$, the field $\chi$ suddenly implodes through the origin.
For the case where $\gamma=0.1$, this pulse interacts with the \qball{}
and induces relatively large oscillations in the scalar field modulus $|\phi|$
which slightly distort the \qball{} profile.
Additionally, the asymmetry of the
imploding pulse imparts a small momentum ``kick" to the \qball{} which
sets it drifting away from the origin very slowly. However, for the case
of $\gamma=0$, the imploding pulse has no effect on the \qball{} and it
remains stationary. By continuing the evolution until $t=1200$, we
observe that these general behaviours continue indefinitely---there is
no significant change to the oscillatory pattern in either case. We
therefore conclude that the corresponding solutions 
are stable.

Turning next to the unstable configurations in
Fig.~\ref{fig:solnspace-log}, we observe two disconnected branches with
distinct behaviour. On the leftmost branch in the figure (labelled
``blowup" and marked by red open circles), we find that the evolutions
quickly become singular as the scalar field grows without bound in
response to the perturbations. 
As described in \cite{Kinach2023}, this behaviour can reasonably be attributed
to the potential \eqref{eqn:log} being unbounded from below.  In particular, it
may become energetically favourable for the scalar field modulus to increase as
the perturbations drive the field to a state of minimal $V(|\phi|)$. However,
since there is no lower bound on $V(|\phi|)$ for large $|\phi|$, the energy
density can become locally negative and the growth can continue indefinitely
in a runaway effect.  Since the resulting configurations do not retain any
resemblance to the initial \qball{}, we classify them as unstable.  We note
that similar behaviour has also been observed in other \qball{} models which
can attain negative energy densities \cite{Anderson1970,Panin2019}.

\begin{figure*}
    \includegraphics[width=\linewidth]{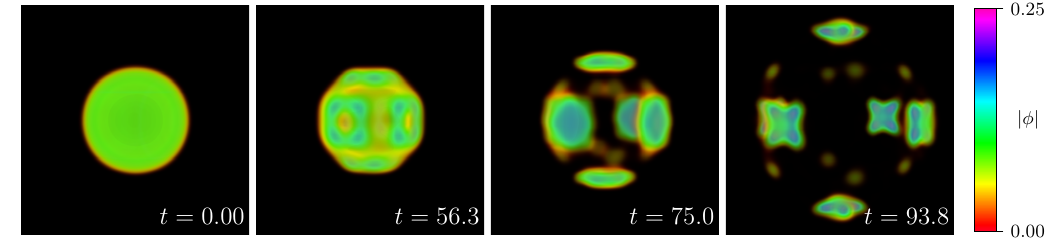}
    \caption{Evolution of the scalar field modulus $|\phi|$ for the
    ``gauged \qshell{}" corresponding to configuration B in Table
    \ref{table:solns}.  A three-dimensional view is shown; at the
    initial time, the fields are shell-like. As the evolution proceeds,
    the shell quickly breaks apart into smaller components which
    propagate away from the origin. Note that we have set $\gamma=0$ for
    this evolution (i.e., the fields are perturbed only by the inherent
    numerical error of the simulation).}
    \label{fig:qshell}
\end{figure*}

On the rightmost unstable branch of Fig.~\ref{fig:solnspace-log} (marked
by red solid circles), we observe that the gauged \qballs{} are quickly
destroyed in response to the perturbations and can evolve in several
ways. The most common outcome is the fragmentation of the original
\qball{} into several smaller components. As an illustrative example, we
plot in Fig.~\ref{fig:qshell} the evolution of a gauged \qball{} which
corresponds to configuration B in Table \ref{table:solns}.  This
configuration is noteworthy in that it represents a shell-like
concentration of the fields (a ``gauged \qshell{}" \cite{Heeck2021shell}) at
the initial time. 
As the evolution proceeds, we observe that the \qshell{} eventually breaks
apart into six main components which travel coincident with the
coordinate axes. We note that this instability, along with every other
instability on the unstable branches of Fig.~\ref{fig:solnspace-log}, can
manifest quickly even without the influence of the perturbing field $\chi$
(i.e., with $\gamma=0$).  However, the specific manner in which the \qball{}
breaks apart
will depend on the configuration under study.

One notable feature of the evolution depicted in Fig.~\ref{fig:qshell} is the
absence of any ring-like structures (``gauged \qrings{}") after the \qshell{}
has broken apart. For the equivalent evolution in axisymmetry (see Fig.~7 of
\cite{Kinach2023}), it has been reported that this particular configuration can
result in the formation of gauged \qrings{} which survive for some time.
However, the absence of such structures in Fig.~\ref{fig:qshell} suggests that
the creation of \qrings{} may be suppressed in full 3D. While we have still
observed the formation of rings in other cases, we find that they are rare and
usually break apart into smaller gauged \qballs{} shortly after they appear.
This indicates that long-lived gauged \qrings{} may be considerably less common
in three spatial dimensions (at least, for the type of evolutions and
perturbations described here).

\begin{figure}
    \includegraphics[width=\columnwidth]{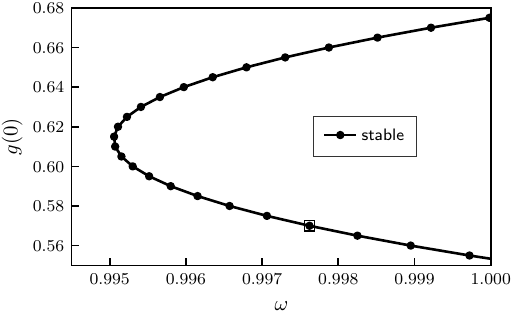}
    \caption{Shooting results and regions of stability for gauged
    \qballs{} in the polynomial model \eqref{eqn:poly} with $e=0.17$.
    Plotted is the numerical shooting parameter $g(0)=\omega-eA_0(0)$
    versus the \qball{} oscillation frequency $\omega$. All
    configurations tested in the model (represented by black solid circles)
    are found to be stable with respect to generic three-dimensional
    perturbations. The open square represents the location of 
    configuration C from Table \ref{table:solns}.}
    \label{fig:solnspace-poly}
\end{figure}

Next, we consider gauged \qball{} stability for
the polynomial potential \eqref{eqn:poly} with $e=0.17$. Once again, we
begin the analysis by applying the shooting procedure of
Sec.~\ref{sec:initdata} to find gauged \qball{} configurations in the
model. The space of solutions for this case is shown in
Fig.~\ref{fig:solnspace-poly}. As stated previously, the choice $e=0.17$
is near the maximum allowable for the polynomial potential and no gauged
\qballs{} can be found with $\omega>1$ \cite{Loginov2020}. 
This significantly limits the range of possible solutions at large gauge
coupling. Similar to the case of the logarithmic model, we evolve each
configuration in Fig.~\ref{fig:solnspace-poly} twice (once with $\gamma=0$ and
once with $\gamma=0.1$) up to at least $t=1200$ in order to assess the stability.
Notably, we find no evidence for configurations which are unstable with respect
to three-dimensional perturbations. This agrees with what has previously been
reported for the equivalent evolutions in axisymmetry \cite{Kinach2023}. 

To conclude this section, let us examine the stability of gauged
\qballs{} for the polynomial potential \eqref{eqn:poly} with $e=0.02$.
In this case, the gauge coupling is much smaller than what has been
considered above and the space of possible solutions is correspondingly
larger. We previously examined this scenario in axisymmetry
\cite{Kinach2023} and found that the transition points between stability
and instability in the solution space match closely with the transition
points predicted for non-gauged \qballs{} with $e=0$.  However, it was
also noted that some solutions near the transition point exhibit ``large
oscillations in the \qball{} interior which significantly disrupt the
shape of the configuration but do not cause the \qball{} to immediately
break apart". Since these solutions could not definitively be said to
retain their initial shape, they were classified as unstable.  Moreover,
the recent results of \cite{Rajaraman2024} suggest a discrepancy between
the transition point predicted by analytical calculations and the
transition point identified numerically in \cite{Kinach2023}. Motivated
by these factors, we now revisit this scenario and examine the
same phenomenon using our fully three-dimensional code.

\begin{figure}
    \includegraphics[width=\columnwidth]{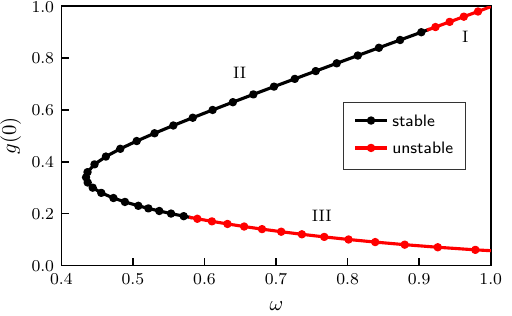}
    \caption{Shooting results and regions of stability and instability for
    gauged \qballs{} in the polynomial model \eqref{eqn:poly} with $e=0.02$.
    Plotted is the numerical shooting parameter $g(0)=\omega-eA_0(0)$ versus
    the \qball{} oscillation frequency $\omega$.  The black solid circles along
    branch II represent configurations which are found to be stable with
    respect to generic three-dimensional perturbations. The red solid circles
    along branches I and III represent configurations which are found to
    be unstable with respect to these perturbations.}
    \label{fig:solnspace-poly_e0.02}
\end{figure}

In Fig.~\ref{fig:solnspace-poly_e0.02}, we plot the space of solutions for
gauged \qballs{} in the polynomial model \eqref{eqn:poly} with $e=0.02$. The
curve can be broken down into three branches: an upper unstable branch I, a
stable branch II, and a lower unstable branch III. Notably, the lower part
of branch II and all of branch III are characterized by scalar field profiles
which are step function-like and resemble the thin-wall \qballs{}
\cite{Heeck2021thinwall}. Once again, we perturb each configuration twice by setting
$\gamma=0$ and $\gamma=0.1$. Any gauged \qballs{} which are clearly
destroyed in response to either perturbation are classified as unstable while
those which oscillate weakly or return toward the original configuration are
classified as stable. For the solutions along branch I, we also observe that the \qballs{}
appear to collapse into solutions which lie along the stable branch II; we also
classify these as unstable, though we comment that this behaviour makes it somewhat difficult to precisely
identify the onset of instability. The salient feature of
Fig.~\ref{fig:solnspace-poly_e0.02} in comparison to Fig.~12 of
\cite{Kinach2023} is the different location for the transition point between
branches II and III of the figure. In particular, this transition point is
found to occur at a larger value of $\omega$ in three spatial dimensions and
the ``large oscillations" observed in axisymmetry are altogether absent. To
verify this claim further, we have evolved the configurations with $g(0)<0.34$ in
Fig.~\ref{fig:solnspace-poly_e0.02} up to at least $t=5000$.
Since the 3D simulations are expected
to fully capture all unstable modes which would arise under axisymmetry
assumptions, we conclude that this is a distinct result from what was reported
in \cite{Kinach2023}.

The origin of the ``large oscillations" observed in axisymmetry is
therefore puzzling, though it might reasonably be attributed to the
unique numerical challenge of evolving the gauged \qballs{} which lie
along the lower part of branch II and branch III. In particular, the
large thin-wall shape of these solutions results in sharp field
gradients arising near the edge of the \qball{}. This can make it
difficult to smoothly resolve the \qball{} boundary unless significant
computational resources are expended. At the same time, we find that the
instabilities of the \qballs{} along this branch may only definitively
manifest after several thousand time steps. This contrasts what is
observed for other unstable gauged \qballs{} in the logarithmic and
polynomial models where the instabilities become obvious rather quickly.
Together, these factors might result in the accumulation of numerical
errors at late times which obscure the stability picture. For example,
the oscillations observed in axisymmetry might possibly be due to a
``de-phasing" of the periodic parts (real and imaginary) of the complex
scalar field which eventually build up and disfigure the \qball{}
profile. However, the fourth-order finite-difference scheme used in the present work
is of a higher accuracy than the second-order method used in \cite{Kinach2023}, so this
may explain why such numerical artefacts are not observed here.
Alternatively, the oscillations observed in axisymmetry may arise due to the
different boundary conditions used or due to problems with
the regularity of the evolved fields along the axis of symmetry at late times.
In any case, the results of Fig.~\ref{fig:solnspace-poly_e0.02} suggest that the location of the
instability threshold for these gauged \qballs{} does not correspond so
nearly with the prediction made by the stability criterion
$(\omega/Q)\,dQ/d\omega<0$ \cite{Correia2001}. This contrasts what was previously reported in
\cite{Kinach2023} but appears to agree with recent analytical findings
\cite{Rajaraman2024}.

\subsection{Collisions}

We now consider relativistic collisions of gauged \qballs{} in three spatial
dimensions.  To construct the binary system, we use the procedure described in
Sec.~\ref{sec:initdata}.  The \qballs{} are initialized at $z=\pm 25$ with 
initial velocities in the range $0.2\leq v \leq 0.8$. Additionally, we define
the impact parameter $b$ as the linear distance between the center of the each
\qball{} in the plane perpendicular to the initial motion. In our
evolutions, we also test the effects of the relative phase difference $\alpha$
and the relative sign of the Noether charge $Q$ on the outcome of the
collision.  The phase difference $\alpha$ is defined through a modification of
the basic \qball{} ansatz \eqref{eqn:sph-ansatz-f},
\begin{equation}
 \label{eqn:sph-ansatz-mod}
  \phi(t,\vec{x})=f(r)\,e^{\epsilon(i \omega t) + i\alpha}.
\end{equation}
By adjusting $\alpha\in[0,\pi]$ for one \qball{} in the binary, a
relative difference in phase can be introduced into the system. This
phase difference is preserved until the moment of impact for collisions
of \qballs{} with identical $\omega$. Additionally, adjusting the
parameter $\epsilon=\pm 1$ (while also taking $A_0(r)\rightarrow-A_0(r)$
in \eqref{eqn:sph-ansatz-A}) for one \qball{} in the binary can flip the
overall sign of its Noether charge $Q$. In this manner, the dynamics of
\qball/anti-\qball{} collisions can be investigated.

For all results presented below, we restrict our analysis to collisions
involving configurations A and C in Table \ref{table:solns}. Since
configuration A is identical to configuration LogC in \cite{Kinach2024},
and since configuration C is identical to configuration PolyB in
\cite{Kinach2024}, this enables a direct comparison between the
collision dynamics in axisymmetry and the equivalent dynamics in three
spatial dimensions.  To facilitate this comparison, we have performed a
number of head-on collision simulations of gauged \qballs{} in 3D; we
find the dynamics of these collisions to be broadly consistent with the
axisymmetric case.  In the discussion below, we will briefly review
these results before turning to collision scenarios with non-zero impact
parameter (which are unique to 3D).

\begin{figure*}
    \includegraphics[width=\linewidth]{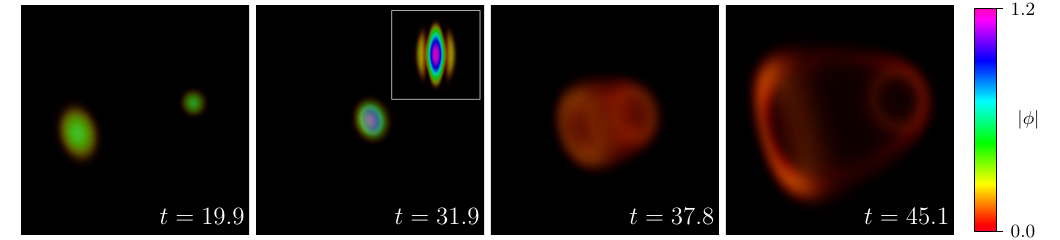}
    \caption{Evolution of the scalar field modulus $|\phi|$ for a collision
    involving configuration A from Table \ref{table:solns} with equal charge,
    velocity $v=0.8$, phase difference $\alpha=0$, and impact parameter $b=0$.
    A three-dimensional off-angle view is shown. The \qballs{} collide at
    $t\approx 32$ and interfere destructively; this is shown from a cross-sectional side-on
    perspective in the inset graphic of the second panel. After the collision,
    the field content predominantly takes the form of two \qrings{} which also
    carry a cylindrical ``wake" of scalar matter.}
    \label{fig:qring}
\end{figure*}

We first discuss the effects of the initial velocity $v$ on the outcome of
head-on collisions with equal charge.  For both A and C in Table
\ref{table:solns}, we find that the Coulomb repulsion of the gauged \qballs{}
can prevent any significant overlap of their respective scalar field content at
low collision velocities. Instead, the \qballs{} travel toward each other,
reach a turning point of vanishing speed and then propagate back toward the
boundaries. This occurs for $v\lesssim 0.3$ for configuration A and $v\lesssim 0.2$ for
configuration C. At higher velocities, the
gauged \qballs{} are able to overcome their mutual repulsion and can behave in
several different ways. For configuration A, we find that the outcome is
typically a fragmentation of the gauged \qballs{} into several smaller
components. In most cases, a significant fraction of each original \qball{}
continues to travel along the $z$-axis after the collision.  This is usually
accompanied by the formation of smaller field remnants which are left behind
near the origin and may travel away in different directions.  For the case of
configuration C, we find that the equivalent collisions result in the merger of
the gauged \qballs{} along with the emission of significant field content in
the form of outgoing waves.
At the highest collision velocities (e.g., $v\gtrsim 0.7$ for configuration A
and configuration C), an increasing fraction of the field content travels
parallel to the $z$-axis after the collision. As illustrated in
Fig.~\ref{fig:qring}, this is accompanied by the development of a destructive
interference pattern in $|\phi|$ at the moment of impact as well as the
formation of gauged \qrings{} in the case of configuration A. 

We now turn to head-on collisions of gauged \qballs{} with phase
differences and opposite charges. It is well-known that the introduction
of a phase difference can induce charge transfer between colliding
\qballs{} \cite{Battye2000}. Here we observe similar behaviour using
$\alpha=\pi/4$ as a sample value. 
As in \cite{Kinach2024}, we find that the gauged \qballs{} created
during the charge transfer process will often fragment into smaller
\qballs{} or even create transient \qrings{}.
In the case of configuration C, we also find some examples where the
gauged \qballs{} created during the collision will almost completely dissipate.
However, the rate of charge transfer is found to
decrease as $v\rightarrow 1$ in both cases. For head-on collisions with
opposite charges, we find that the Coulomb force (which is now
attractive) can accelerate the gauged \qballs{} prior to the moment of
impact. After the collision, the total Noether charge in the system is
reduced as the \qballs{} have partially annihilated. This process can
create smaller \qball{} remnants which lag the main \qballs{} (which are
now highly perturbed) and propagate along or away from the $z$-axis. It
can also produce a wake of scalar radiation or a 
quasispherical pulse of
electromagnetic radiation which emanates from the origin. In general, we
find that the amount of charge which is annihilated depends on
the collision velocity, with the least amount of annihilation occurring
at the largest velocities.

While the above results are broadly consistent with the equivalent
calculations in axisymmetry \cite{Kinach2024}, we comment here on some
subtle differences. One main difference relates to the behaviour of any
gauged \qrings{} which are created during the collisions.  In
axisymmetry, \qrings{} were found to be a rather common outcome of
intermediate- and high-velocity collisions that resulted in gauged
\qball{} fragmentation.  In these cases, the rings tended to propagate
some distance away from the origin before collapsing back onto the axis
of symmetry at late times (though this final fate could not be confirmed
in all cases).  While we have still observed the formation of gauged
\qrings{} in our fully three-dimensional simulations, we find that they
tend to quickly break apart into a number of spherical gauged \qballs{}
in the majority of cases. It is only in rare circumstances (such as the
scenario depicted in Fig.~\ref{fig:qring}) where we have observed that
the \qrings{} can survive long enough to reach a radius which is many
times greater than the size of the original \qball{}. This reaffirms
our comments in Sec.~\ref{sec:stability} that \qring{} formation, while
not explicitly forbidden, may be a rare phenomenon in the absence of
symmetry restrictions. 

\begin{figure*}
    \includegraphics[width=\linewidth]{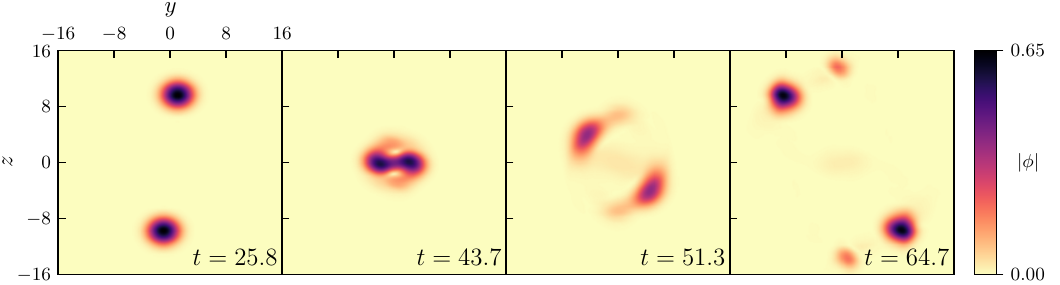}
    \caption{Evolution of the scalar field modulus $|\phi|$ for a
    collision involving configuration A from Table \ref{table:solns}
    with equal charge, velocity $v=0.6$, phase difference $\alpha=0$,
    and impact parameter $b=2$. A two-dimensional slice through the
    $x=0$ plane is shown. The \qballs{} collide at $t\approx 43$ and
    fragment into smaller components which travel away in different
    directions. While the dynamics in this case are mostly planar, we
    comment that small amounts of field content also propagate away from the
    collision plane; this field content is not shown in the figure.}
    \label{fig:scatter}
\end{figure*}

Having discussed the dynamics of head-on collisions, we now focus on the
case where the impact parameter $b$ is non-zero. Since these ``off-axis"
collisions are obviously forbidden in axisymmetry, they represent a
novel dynamical scenario which has not been explored in the previous
studies. We begin by considering off-axis collisions of equal-charge
gauged \qballs{}. In this case, we find that a common outcome is the
``deflection" of the gauged \qballs{} due to the influence of the
repulsive gauge field.  This can result in the \qballs{} following a
discernible curved trajectory which makes an angle $\theta$ with the
$z$-axis at late times.
The exact value of $\theta$ for a given collision can depend on several
factors such as
the initial velocity $v$ and the impact parameter $b$.
For equal-charge collisions, we find that $\theta$ is generally maximized when $v$ and $b$ are small (in
fact, one could interpret the repulsive scenario discussed above for head-on
collisions with equal charge and low velocity as a case of maximal
deflection where $\theta=\pi$). However, when $v$ is sufficiently large
and $b$ is not larger than the approximate \qball{} width, the scalar fields
from each \qball{} can ``graze" each other during the collision. In this case,
the end result may be a fragmentation or merger of the gauged \qballs{}. In
Fig.~\ref{fig:scatter}, we plot a ``grazing" collision of configuration A from
Table \ref{table:solns} with equal charge, velocity $v=0.6$, phase difference
$\alpha=0$, and impact parameter $b=2$. The gauged \qballs{} collide at
$t\approx43$ with a majority of the field content emerging at an angle $\theta
\approx \pi/4$ with respect to the $z$-axis.  We also observe that the initial
gauged \qballs{} have partially fragmented into smaller objects which travel
close to the $z$-axis.  Repeating the calculation
shown in Fig.~\ref{fig:scatter} for a variety of choices of $v$ and $b$, we
find that the outcomes are broadly consistent with what has been described
above, though the deflection angles and fragmentation products may differ
depending on the specific collision parameters.

\begin{figure*}
    \includegraphics[width=\linewidth]{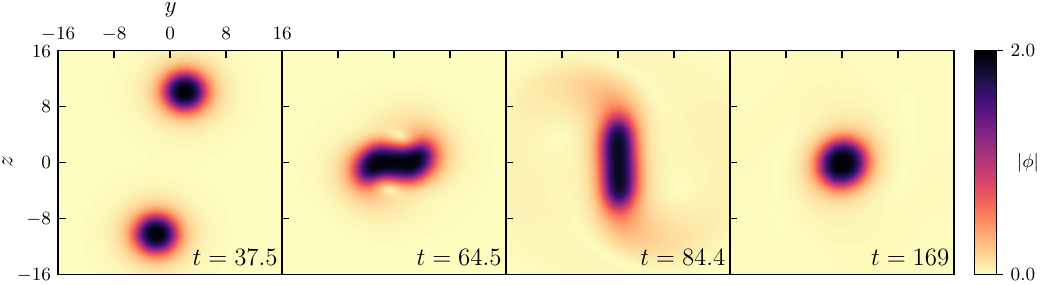}
    \caption{Evolution of the scalar field modulus $|\phi|$ for a collision involving configuration
    C from Table \ref{table:solns} with equal charge, velocity $v=0.4$,
    phase difference $\alpha=0$, and impact parameter $b=4$. A two-dimensional slice through the $x=0$ plane
    is shown. The \qballs{} collide at $t\approx 64$ and merge into a single gauged \qball{} which remains 
    at the origin. In this process, a considerable amount of the field content is radiated away toward the
    boundaries.} 
    \label{fig:merger}
\end{figure*}

In Fig.~\ref{fig:merger}, we plot a collision involving configuration C
from Table \ref{table:solns} with equal charge, velocity $v=0.4$, phase
difference $\alpha=0$, and impact parameter $b=4$. In contrast to what
is shown in Fig.~\ref{fig:scatter} for configuration A, here we see that
the end result is a merger of the original gauged \qballs{}.  During the
merger process, a significant amount of field content is radiated away
toward the boundaries in the form of aspherical waves. By $t\approx 169$
(the last panel in the figure), the merged configuration has settled
down into a single gauged \qball{} centered at the origin which remains 
slightly perturbed.
The properties of this final merged state turn out to be
similar in some ways to the properties of configuration C before the
collision.  For example, the scalar field attains a value of
$|\phi|\approx 1.98$ at the origin by $t\approx 169$ while the
oscillation frequency (which we determine by tracking the real part of
the scalar field during the collision) is found to be
$\omega\approx0.99$ in the merged state. This result might be
expected for gauged \qballs{} with $e=0.17$ in the potential
\eqref{eqn:poly} since the space of possible solutions is extremely
small (see Fig.~\ref{fig:solnspace-poly}). For configuration C, we find
that mergers are a common outcome for moderate values of the collision
velocity and impact parameter. At larger values of $v$ and $b$, the
gauged \qballs{} can avoid the merged state through (for example) deflection
of the fields.

It is worthwhile to discuss the final state of Fig.~\ref{fig:merger} in
greater detail. Due to the off-axis motion of the binary, the total
angular momentum of the system is non-zero at the initial time. 
It is plausible that some of this angular momentum may be retained by the
merged configuration at late times, potentially representing an object
analogous to a spinning \qball{} \cite{Volkov2002,Campanelli2009}. At a
visual level, the elongated and ``rotating" appearance of $|\phi|$ in
the second and third panel of Fig.~\ref{fig:merger} may also seem to
support this idea.  However, there are several reasons why the final
merged state is unlikely to represent a configuration of this type.
First, we observe that the gauged \qball{} very quickly returns to a
near-spherical shape by $t\approx 169$ through the emission of
significant field content toward the boundaries. 
However, field
configurations with angular momentum are not expected to be
spherically-symmetric and may also be characterized by the presence of
nodes away from the center \cite{Volkov2002}. Second, we have explicitly
computed the angular momentum tensor,
\begin{equation}
  M^{ij}=\int(x^iT^{j0}-x^jT^{i0})\,d^3x,
\end{equation}
and found that the $x$-component of the angular momentum, $J_x=M^{23}$,
is almost totally radiated away from the origin by $t\approx 169$. Since the
angular momentum of a spinning \qball{} (at least, in the non-gauged case) is
expected to be an integer multiple of the Noether charge $Q$, we conclude that
mergers of this type are unlikely to represent the usual spinning structures.
At the same time, we cannot rule out the possibility that some small amount of
angular momentum will still be retained in the merged state even at later
times. If so, the configuration might be analogous to the ``slowly rotating"
\qballs{} recently proposed in \cite{Almumin2024}.

\begin{figure*}
  \includegraphics[width=\linewidth]{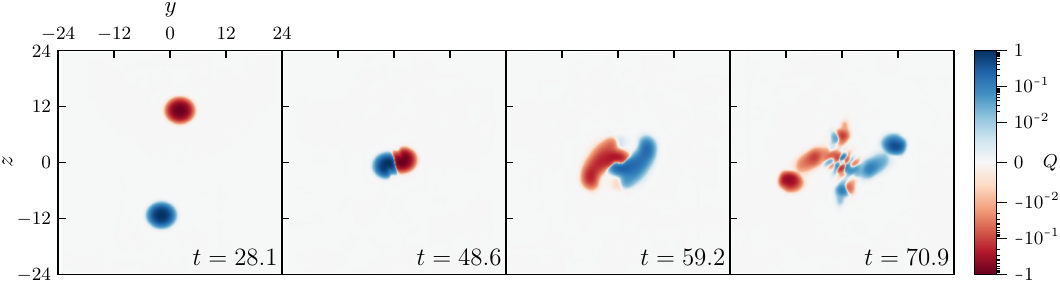}
    \caption{Evolution of the Noether charge $Q$ for a collision
    involving configuration A from Table \ref{table:solns} with opposite
    charge, velocity $v=0.5$, phase difference $\alpha=0$, and impact
    parameter $b=4$. A two-dimensional slice through the $x=0$ plane is
    shown. The \qballs{} collide at $t\approx 48$ and fragment into
    smaller components after partially annihilating.  While the dynamics
    in this case are mostly planar, we comment that small portions of
    charge also propagate away from the
    collision plane; these small charges are not shown in the figure.
    Note that a hybrid colormap is used: charge values below
    $|Q|=10^{-2}$ are mapped linearly to zero while values above this
    threshold are mapped logarithmically to the charge maximum.  }
    \label{fig:annihilation}
\end{figure*}

\begin{figure*}
    \includegraphics[width=\linewidth]{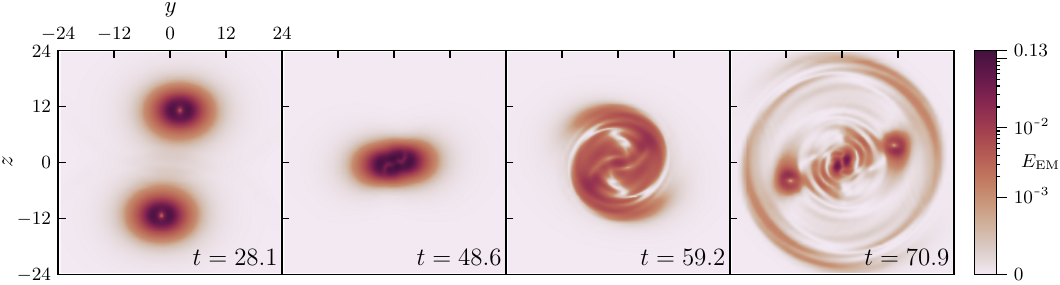}
    \caption{Evolution of the electromagnetic field energy $E_{\rm EM}$
    for a collision involving configuration A from Table
    \ref{table:solns} with opposite charge, velocity $v=0.5$, phase
    difference $\alpha=0$, and impact parameter $b=4$. A two-dimensional
    slice through the $x=0$ plane is shown. The \qballs{} collide at
    $t\approx 48$ and fragment into smaller components after partially
    annihilating.  After the collision, a pulse of electromagnetic
    energy emanates from the origin (fourth panel).  The shape of this
    pulse is not limited to the $y$--$z$ plane shown here; it can be
    seen to propagate in all directions when viewed three-dimensionally.
    Note that a hybrid colormap is used: energy values below $E_{\rm
    EM}=10^{-3}$ are mapped linearly to zero while values above this
    threshold are mapped logarithmically to the energy maximum.}
    \label{fig:radiation}
\end{figure*}

Next we turn to off-axis collisions of gauged \qballs{} with opposite
charges. Unlike the repulsive behaviour seen for the equivalent
collisions with equal charge, here we observe that the \qballs{}
experience an attractive acceleration which curves their trajectories
toward the origin. If the impact parameter and initial velocity are
sufficiently large, the \qballs{} may pass by one another without any
significant interaction between their respective scalar fields.  This is
similar to the ``deflection" described above for the equal-charge
collisions, though now the deflection occurs in the opposite direction
(i.e., toward the other \qball{} in the binary rather than away from
it).  If the impact parameter is small, the \qballs{} will generally
experience a ``grazing" collision which can result in several possible
outcomes. Most commonly, the gauged \qballs{} will partially annihilate
and fragment into a number of smaller components (for the case of
configuration A) or radiate a portion of the field content toward the
boundaries (for the case of configuration C); this is similar to their
behaviour during head-on collisions.  In Fig.~\ref{fig:annihilation}, we
plot the Noether charge density $Q$ for a grazing collision involving
configuration A from Table \ref{table:solns} with initial velocity
$v=0.5$, phase difference $\alpha=0$, and impact parameter $b=4$. During
the collision, the \qballs{} complete a partial orbit around each other
before escaping along a trajectory which is roughly perpendicular to
their initial motion. A number of positively- and negatively-charged
remnants are also created during the collision in the vicinity of the
origin. By $t\approx 70.9$, approximately half of the total charge
in the system has been annihilated. The acceleration and annihilation of
charges during this process can also result in the production of an
electromagnetic radiation pulse. In Fig.~\ref{fig:radiation}, we plot
the energy contained in the electromagnetic field,
\begin{equation}
  E_{\rm EM}=\frac{1}{2}\left(|\vec{E}|^2+|\vec{B}|^2\right),
\end{equation}
where $\vec{E}$ and $\vec{B}$ are the electric and magnetic field
vectors, respectively. By comparing Fig.~\ref{fig:annihilation} and
Fig.~\ref{fig:radiation}, we can see that a pulse of outgoing energy is
created in the electromagnetic field which does not correspond to any
significant amount of charge. We interpret this as representing
electromagnetic radiation. We find the production of electromagnetic
radiation to be a general phenomenon associated with gauged
\qball{}/anti-\qball{} collisions, though the exact amount of radiation
produced may depend on both the motion of the charges and the total
amount of annihilation which occurs in the system.


To conclude this section, let us comment briefly on the off-axis
collision of gauged \qballs{} with a phase difference of $\alpha=\pi/4$.
Similar to the case of head-on collisions, we find that the introduction
of a relative phase difference can result in the transfer of charge
between the colliding \qballs{}. When the impact parameter is non-zero,
the dynamics of this charge transfer can be altered in minor ways. For
example, the charge transfer may occur asymmetrically such that the
resulting \qballs{} are left travelling at an angle relative to their
initial motion; this angle can depend on both the collision velocity and
the impact parameter. As the impact parameter is further increased, the amount
of charge transfer appears to be reduced due to the smaller surface of contact
between the colliding \qballs{}.
Otherwise, the charge
transfer during off-axis collisions can generally be said to resemble
the results for head-on collisions (including phenomena such as
fragmentation or dissipation of the resulting \qballs{}).

\section{\label{sec:conclusion} Conclusion }

In this work, we have studied the dynamical behaviour of $U(1)$ gauged
\qballs{} using fully three-dimensional numerical evolutions. First, we
investigated the classical stability of gauged \qballs{} with respect to
generic three-dimensional perturbations. Second, we explored the
dynamics of gauged \qballs{} during head-on and off-axis collisions
at relativistic velocities.

With regard to stability, we have found numerical evidence for gauged
\qballs{} which remain stable against generic perturbations over long
dynamical timescales. To reach this conclusion, we have perturbed the
\qballs{} in two different ways: through the inherent numerical error of
our finite-difference implementation and through the interaction of an
auxiliary scalar field which acts as a perturbing agent. Testing
configurations in the logarithmic model, we have found evidence for both
stable and unstable branches in the solution space. The solutions on the
stable branch tend to respond to the perturbations by oscillating
continuously near the initial configuration. The solutions on the
unstable branch are found to break apart in various ways (usually into a
number of smaller gauged \qballs{}). We have also tested configurations
in the sixth-order polynomial scalar field model, finding no evidence
of unstable configurations for our choice of the model parameters with $e=0.17$.
Finally, we have revisited the case of $e=0.02$ in the polynomial model and
found a new result for the transition point between stability and instability
in the solution space. This result differs from what was found in \cite{Kinach2023}
but appears to be in agreement with recent analytical findings \cite{Rajaraman2024}.

With regard to relativistic collisions of gauged \qballs{}, we have
tested the effect of the initial velocity, relative phase, relative
charge, and impact parameter on the outcome of the collision. For the
case of head-on collisions, we have found that the dynamics in three
spatial dimensions are broadly consistent with previous results reported
under axisymmetry assumptions \cite{Kinach2024}.  For the case of
off-axis collisions, we have found that the impact parameter can play a
significant role in modifying the collision outcome. For example, the
gauged \qballs{} can experience attractive or repulsive ``deflections"
from their initial trajectories depending on their relative charges,
velocities, and the collision impact parameter.  In other cases, the
\qballs{} may experience ``grazing" collisions which can modify the
dynamics during \qball{} fragmentation and mergers. Aside from these
differences, the main phenomena associated with these collisions (such
as charge transfer, annihilation, and radiation production) are found to
be similar to the head-on case.

The results of this work are significant for several reasons.  First,
they address the general question of gauged \qball{} classical stability
which was originally raised in \cite{Panin2017}. Second, they provide
new insights into the time-dependent behaviour of gauged \qballs{} in
realistic three-dimensional settings.  Together, these results may be
relevant for future studies of \qballs{} in various physical contexts
(such as in early-Universe cosmology). At the same time, we hope that
this work may inspire further numerical explorations of related soliton
models such as Proca \qballs{} \cite{Heeck2021proca}, spinning \qballs{}
\cite{Volkov2002,Campanelli2009,Almumin2024}, and charge-swapping
\qballs{} \cite{Copeland2014,Xie2021,Hou2022}.

\acknowledgments

This work was supported by the Natural Sciences and Engineering Research
Council of Canada. Computing resources were provided by the Digital Research
Alliance of Canada and the University of British Columbia.\\

\appendix

\section{Evolution Equations in Three Spatial Dimensions}
\label{sec:evoeqns}
When expressed using the coordinates defined by \eqref{eqn:minkowski},
the evolution equations for the system
\eqref{eqn:eom-a}--\eqref{eqn:eom-b} take on the following form:
\begingroup
\allowdisplaybreaks
\begin{widetext}
\begin{align}
  \partial_t^2\phi_1=&\;\partial_x^2 \phi_1 + \partial_y^2 \phi_1 + \partial_z^2 \phi_1 + 2e\left( -A_t \partial_t\phi_2 +A_x\partial_x\phi_2 + A_y\partial_y \phi_2 +A_z\partial_z \phi_2\right) -e^2\phi_1 \left( -A_t^2 +A_x^2 + A_y^2 + A_z^2 \right)-\frac{1}{2}\partial_{\phi_1}V(\phi_1,\phi_2), \label{eqn:phi1res}\\
  \partial_t^2\phi_2=&\;\partial_x^2 \phi_2 + \partial_y^2 \phi_2 + \partial_z^2 \phi_2 - 2e\left( -A_t \partial_t\phi_1 +A_x\partial_x\phi_1 + A_y\partial_y \phi_1 +A_z\partial_z \phi_1\right) -e^2\phi_2 \left( -A_t^2 +A_x^2 + A_y^2 + A_z^2 \right)-\frac{1}{2}\partial_{\phi_2}V(\phi_1,\phi_2), \label{eqn:phi2res}\\
  \partial_t^2 A_t=&\;\partial_x^2 A_t + \partial_y^2 A_t + \partial_z^2 A_t + 2e\left( \phi_1\partial_t\phi_2 - \phi_2\partial_t\phi_1 \right) - 2e^2 \left( \phi_1^2+\phi_2^2\right) A_t, \\[2pt]
  \partial_t^2 A_x=&\;\partial_x^2 A_x + \partial_y^2 A_x + \partial_z^2 A_x + 2e\left( \phi_1\partial_x\phi_2 - \phi_2\partial_x\phi_1 \right) - 2e^2 \left( \phi_1^2+\phi_2^2\right) A_x, \\[2pt]
  \partial_t^2 A_y=&\;\partial_x^2 A_y + \partial_y^2 A_y + \partial_z^2 A_y + 2e\left( \phi_1\partial_y\phi_2 - \phi_2\partial_y\phi_1 \right) - 2e^2 \left( \phi_1^2+\phi_2^2\right) A_y, \\[2pt]
  \partial_t^2 A_z=&\;\partial_x^2 A_z + \partial_y^2 A_z + \partial_z^2 A_z + 2e\left( \phi_1\partial_z\phi_2 - \phi_2\partial_z\phi_1 \right) - 2e^2 \left( \phi_1^2+\phi_2^2\right) A_z. \label{eqn:Azres}
\end{align}
\end{widetext}
\endgroup
\noindent
Here, the
subscripts $\{t,x,y,z\}$ correspond to the spacetime coordinates while the
subscripts $\{1,2\}$ denote the real and imaginary parts of the scalar
field, respectively. In deriving \eqref{eqn:phi1res}--\eqref{eqn:Azres},
we have invoked the Lorenz gauge condition \eqref{eqn:lorenz} as a means
to simplify the equations. After applying the coordinate transformations
\eqref{eqn:compactX}--\eqref{eqn:compactZ}, we solve these equations
using the fourth-order finite-difference scheme described in
Sec.~\ref{sec:evoscheme} together with the initial data procedure of
Sec.~\ref{sec:initdata}. 

\section{Code Validation}
\label{sec:codevalid}

\begin{figure}
    \includegraphics[width=\columnwidth]{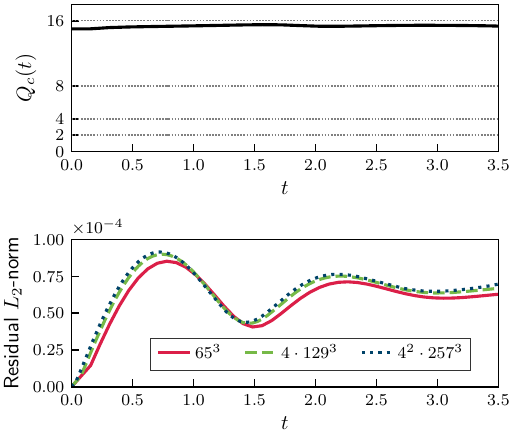}
    \caption{Representative results for a three-level convergence test
    (top panel) and independent residual test (bottom panel) of the
    finite-difference implementation described in
    Sec.~\ref{sec:evoscheme}. In the top panel, the convergence factor
    $Q_c(t)$ is computed for the evolved variable $\phi_1$.
    In the bottom panel, the $L_2$-norm for the independent residual of equation
    \eqref{eqn:phi1res} is computed at grid resolutions of $65^3$,
    $129^3$ and $257^3$. In both cases, the implementation is found
    to be convergent at the expected order.}
    \label{fig:CVtest}
\end{figure}

In order to assess the validity of our code, we have performed a series
of numerical tests of convergence. In these tests, we use generic
Gaussian-like initial data which approximately satisfies the constraint
equations \eqref{eqn:divE}--\eqref{eqn:divB} at the initial time.  We
evolve the data on a uniform grid at various resolutions and compute the
convergence factor $Q_c(t)$ as
\begin{equation}
  Q_c(t)=\frac{\|u^{4h}-u^{2h}\|}{\|u^{2h}-u^{h}\|}.
  \label{eqn:CVfactor}
\end{equation}
Here, $h$ represents the spacing between grid points, $u^n$ represents
the solution computed with grid spacing $n$, and $\|\cdot\|$ denotes the
$L_2$-norm. For a finite-difference scheme with $O(h^m)$ accuracy, one
expects to find $Q_c(t)\rightarrow 2^m$ as $h\rightarrow 0$
\cite{Choptuik2006}. We therefore expect to observe $Q_c(t)\approx 16$
for the fourth-order finite-difference scheme described in
Sec.~\ref{sec:evoscheme}. In the top panel of Fig.~\ref{fig:CVtest}, we
plot the results of this test for the real part of the scalar field,
$\phi_1$, computed in the polynomial model \eqref{eqn:poly} with
$e=0.5$, $h=0.2$, and $m=k=1$. Using grid resolutions of $65^3$,
$129^3$, and $257^3$ to compute $Q_c(t)$ in \eqref{eqn:CVfactor}, we
find that the implementation is convergent to approximately
fourth-order, as we expect. In addition to $\phi_1$, we have also
repeated this test for all other evolved quantities in the equations of
motion \eqref{eqn:phi1res}--\eqref{eqn:Azres}. 
We find similar fourth-order behaviour in each case.

As a secondary test, we have performed an independent residual
evaluation \cite{Choptuik2006} to verify that our numerical solution
reasonably approximates the continuum solution of the problem. In this
test, the solution obtained using our fourth-order finite-difference
scheme is substituted into a separate second-order centered
discretization of the equations of motion
\eqref{eqn:phi1res}--\eqref{eqn:Azres}. If the residuals of these
equations converge away at second-order in the grid spacing
(corresponding to rescaling by factors of four), we conclude that the
original finite-difference scheme has been correctly implemented. The
results of this test are shown in the bottom panel of
Fig.~\ref{fig:CVtest}. Once again, we use grid resolutions of $65^3$,
$129^3$, and $257^3$ and pick equation \eqref{eqn:phi1res} as a
representative example. In the figure, we observe the expected convergence
of the residual at second-order; the residuals for the other evolution
equations \eqref{eqn:phi2res}--\eqref{eqn:Azres} are found to behave in
a similar way. This provides an additional check of the validity of our
implementation.

\bibliographystyle{apsrev4-2}
\bibliography{refs}  

\end{document}